\documentclass[pre,twocolumn,floatfix,aps,showpacs]{revtex4}
\usepackage{amssymb,amsmath,epsfig,graphicx}
\usepackage[latin1]{inputenc}
\usepackage{dcolumn}% Align table columns on decimal point
\usepackage{amsfonts}

\newcommand{\be}{\begin{equation}}
\newcommand{\ee}{\end{equation}}
\newcommand{\bea}{\begin{eqnarray}}
\newcommand{\eea}{\end{eqnarray}}
\newcommand{\bi}{\begin{itemize}}
\newcommand{\ei}{\end{itemize}}

\newcommand{\Fphi}{{\cal F}(\{\varphi\})}

\begin{document}

\title{Noise-induced phase transitions in field-dependent relaxational
dynamics:  The Gaussian ansatz}
\author{Francisco J. Cao$^{1,2}$}
\email{francao@fis.ucm.es}
\author{Kevin Wood$^{3,4}$}
\email{kwood@ucsd.edu}
\author{Katja Lindenberg$^{3}$}
\email{klindenberg@ucsd.edu}

\affiliation{
$^{(1)}$ Departamento de F\'{\i}sica At\'omica, Molecular y
Nuclear, Universidad Complutense de Madrid, Avenida Complutense
s/n,
E-28040 Madrid, Spain\\
$^{(2)}$ LERMA, Observatoire de Paris, Laboratoire Associ\'e au
CNRS UMR
8112, 61, Avenue de l'Observatoire, 75014 Paris, France\\
$^{(3)}$Department of Chemistry and Biochemistry and Institute for
Nonlinear Science, University of California San Diego, 9500 Gilman
Drive,
La Jolla, CA 92093-0340, USA\\
$^{(4)}$ Department of Physics, University of California San
Diego, 9500 Gilman Drive, La Jolla, CA 92093-0340, USA}

\date{\today}

\begin{abstract}
We present an analytic mean field theory for relaxational dynamics in
spatially extended systems that undergo purely noise-induced phase
transitions to ordered states.  The theory augments the usual mean field
approach with a Gaussian ansatz that yields quantitatively accurate
results for strong coupling. We obtain analytic results not only
for steady state mean fields and distribution widths, but also for the
dynamical approach to a steady state or to collective oscillatory
behaviors in multi-field systems.  Because the theory yields dynamical
information, it can also predict the initial-condition-dependent final
state (disordered state, steady or oscillatory ordered state)
in multistable arrays.
\end{abstract}


\maketitle

\section{Introduction}
The interplay of stochasticity and nonlinearity often leads to a
wide range of interesting and often counterintuitive behaviors.
Noise and nonlinearities have been implicated as sources of both
zero-dimensional transitions~\cite{horsthemke} as well as
nonequilibrium phase transitions in spatially extended
systems~\cite{garcia}. Particularly intriguing among these is a
broad class of macroscopic ordering phenomena that have been
called ``purely-noise-induced" because order is exhibited only in
the presence of noise. These include order-disorder
transitions~\cite{chris,ibanes,buceta04}, the emergence of
periodic spatial structures and pattern formation in increasingly
noisy systems~\cite{wood06,wio03}, and, more recently, purely
noise-induced collective oscillatory
behavior~\cite{wio04,kawai,buceta06}. Purely noise-induced phase
transitions to oscillatory behavior may be dynamically induced and
dependent on the Stratonovich drift, i.e., on the interpretation
of the noise~\cite{chris,kawai}, or it may occur in systems with
field dependent relaxational dynamics where the interpretation of
the noise at most shifts the transition
points~\cite{ibanes,buceta04,wood06,buceta06}. The particular
novelty of the latter transition lies in the fact that it arises
entirely from an energy functional-like relaxational dynamics, and
not from short-time dynamic instabilities which become
strengthened and sustained by spatial coupling.  As such, these
models represent systems that, while decidedly nonequilibrium and
potentially time-dependent in nature, nevertheless maintain
certain parallels with equilibrium statistical models, namely the
evolution towards the minimum of a (possibly time-dependent)
effective potential. Our focus in this paper is on phase
transitions in systems with field dependent relaxational dynamics.

Analytic work on these models has been fairly limited, with most of
the information coming from numerical simulations.  We are not aware of
solvable analytic theories for locally coupled (e.g. nearest neighbor)
noisy extended systems in which any sort of ordering transition is
observed as the system or noise parameters are varied.  Globally coupled
arrays have been more amenable to mean field
theories based on the analytic solution of a
Fokker-Planck equation in the stationary state. This approach has
been successfully applied to
systems that achieve a time-independent steady state.  Such mean field
theories provide
information about the nature of the steady state and the conditions that
lead to disordered vs ordered states.
Thus, while quite useful for ascertaining asymptotic properties,
these static theories provide no insight into the dynamical
evolution toward a steady state.  Such insight is particularly important
in multistable systems.

Most strikingly, these theories fail to provide even a reasonable
asymptotic description of phase transitions to collective
time-dependent behavior, when the complexity of the corresponding
Fokker-Planck equation prohibits a self-consistent analytical
treatment~\cite{buceta06} even in globally coupled arrays. The
complicated behavior seen, for example, in the transition to
noise-induced limit cycles thus requires different methods for
analytic study~\cite{kawai}.

Our purpose here is to provide an approximate analytical method
supplementary to the mean field approach for studying the dynamics
of noise-induced phase transitions in relaxational systems
inspired by a related quantum field method \cite{qf}.  By
introducing a Gaussian ansatz and a series expansion about the
(time-dependent) mean field values, we develop a time-dependent
theory which captures the dynamics of these systems in the limit
of strong spatial coupling.  As well as providing new analytical
insight into the evolution to the steady state in one-field
systems, this treatment yields a set of simple approximate
ordinary differential equations detailing the oscillatory dynamics
and the approach to these dynamics in two-field systems. We note
that an alternative approach to the one adopted in this paper is
to generalize the procedure developed in \cite{kawai} that gives a
set of differential equations for the central moments. It can be
shown that this alternative approach gives the same results at
leading order for large coupling as the Gaussian ansatz, we are
preparing a manuscript where we show the details of the
comparison.

The paper is organized as follows.  In Sec.~\ref{sec2} we briefly recall
the one- and two-field relaxational models.  The mean field equations
and the Gaussian ansatz for their solution are presented in
Sec.~\ref{sec3}. In Sec.~\ref{sec4} we analyze the one-field system and
test our Gaussian ansatz results in the steady state and in the
dynamical approach to the steady state, including regimes of
multistability.  The two-field system is examined in Sec.~\ref{sec5},
where we show that our theory captures the collective oscillatory
behavior, including regimes of multistability.  In both Secs.~\ref{sec4}
and \ref{sec5} we explore approximate solutions to the differential
equations obtained with the Gaussian ansatz to successfully reduce the
problems to quadrature in some regimes.  We conclude with a short
summary in Sec.~\ref{sec6}.

\section{The models}
\label{sec2}

We briefly recall the one-field and two-field relaxational models.
In the one-field case we introduced the Langevin
equation defined on a lattice~\cite{buceta04},
\begin{equation}
\dot\varphi_i = - \Gamma(\varphi_i)
\frac{\delta\Fphi}{\delta\varphi_i} + [\Gamma(\varphi_i)]^{1/2}
\xi_i(t),
\label{langevinonefield}
\end{equation}
where $ \varphi_i $ is the value of the scalar field at site $i$,
$ \Gamma(\varphi_i) $ is the field-dependent kinetic coefficient,
$ \Fphi $ is an energy functional, and $ \xi_i(t) $ is a
spatio-temporal white noise with zero mean and intensity $
\sigma^2 $, $ \langle \xi_i(t) \xi_j(t') \rangle = \sigma^2
\delta_{ij} \delta(t-t') $. All the quantities introduced
previously and those that we introduce later in this paper are
consistently adimensionalized. As the choice of noise
interpretation does not qualitatively affect the
dynamics~\cite{buceta04,wood06,buceta06}, we choose the It\^o
interpretation for convenience. The energy functional includes
local potentials $V(\varphi_i)$ and a simple harmonic coupling
between sites~\cite{buceta04,buceta06}:
\begin{equation}
{\cal F}(\{\varphi\}) = \sum_{i=1}^N \left[ V(\varphi_i) +
\frac{K}{4n} \sum_{\langle ij \rangle} (\varphi_j-\varphi_i)^2
\right].
\label{energyfunctional}
\end{equation}
Here $N$ is the number of lattice sites and $K$ is the coupling
strength.  The sum $ \sum_{\langle ij \rangle} $ runs over
all $n$ sites $j$ coupled to site $i$.  For nearest neighbor coupling
$n=2d$ while for global coupling $n=N-1$ (other forms of coupling
were also considered~\cite{wood06} but
will not be dealt with here).  In~\cite{buceta04} we carried out
comprehensive studies of the phase space diagram in $(K,\sigma^2)$ space
where different disordered and ordered phases can be found for different
types of local potentials $V(\varphi)$ and
of kinetic coefficients $\Gamma(\varphi)$.  In general, necessary
conditions on these functions for the observation of order-disorder phase
transitions include that, at increasingly large $|\varphi|$,
$V(\varphi)$ grow at least harmonically, and $\Gamma(\varphi)$ go
to zero so that the effect of the noise is weakened.  We were able to
analyze the steady states for these systems analytically within
a mean field approximation.  This approximation leads to a Fokker-Planck
equation that can be solved in the steady state, which
allowed us to establish the
phase space diagram for the steady state, but provided no dynamical
information.

The addition of a second field to this problem was introduced
in~\cite{buceta06},
\begin{equation}
\begin{split}
&\dot\varphi_i = - \Gamma(\varphi_i)
\frac{\delta\Fphi}{\delta\varphi_i} +
  [\Gamma(\varphi_i)]^{1/2} \xi_i(t) - \omega z_i, \\
&\dot z_i = \omega \varphi_i,
\end{split}
\label{langevintwofields}
\end{equation}
where $\omega $ is a frequency.  This system undergoes a
noise-induced phase transition to collective oscillatory behavior
when the noise exceeds a critical intensity.  The mean field
Fokker-Planck equation for this two-field system can not be solved
analytically because it remains time dependent for all time. Thus, all
the information obtained up to now has been numerical.

Since our Gaussian ansatz method relies on the mean field evolution
equations, in the next section we briefly review the mean field approach
and introduce the ansatz.

\section{The mean field and the Gaussian ansatz}
\label{sec3}

In the mean field approximation the sum over neighbors connected to site
$i$ in the derivative of the energy functional appearing in
Eq.~(\ref{langevinonefield})
or Eq.~(\ref{langevintwofields}),
\begin{equation}
\frac{\delta{\cal F }(\varphi_i)} {\delta\varphi_i} =
\frac{\partial V(\varphi_i)}{\partial\varphi_i} + K
\left(\varphi_i-\frac{1}{n}\sum_{\langle ij \rangle} \varphi_j
  \right),
\end{equation}
is replaced by the mean field value,
\begin{equation}
\frac{1}{n}\sum_{\langle ij \rangle}\varphi_j(t) \longrightarrow \langle
\varphi(t) \rangle \equiv \varphi_0(t).
\end{equation}
Since all the sites are then equivalent, the lattice index can be
dropped and the set of field equations reduces to a single equation
(one-field system),
\begin{equation}
\dot{\varphi}(t) = a(\varphi; \varphi_0(t))
+\left[\Gamma(\varphi)\right]^{\frac{1}{2}}\xi(t),
\label{meanfieldone}
\end{equation}
or to two coupled equations (two-field system),
\begin{equation}
\begin{split}
&\dot{\varphi}(t) = a(\varphi; \varphi_0(t))
+\left[\Gamma(\varphi)\right]^{\frac{1}{2}}\xi(t) -
\omega z,\\
&\dot{z} = \omega \varphi,
\end{split}
\label{meanfieldtwo}
\end{equation}
where
\begin{equation}
a(\varphi; \varphi_0(t)) \equiv - \Gamma(\varphi) \left\{\frac{
\partial V(\varphi)}{\partial \varphi}-K[\varphi_0(t)
-\varphi] \right\}.
\end{equation}
The unknown mean field $\varphi_0(t)$ must be determined
self-consistently,
\begin{equation}
\varphi_0(t)=\langle \varphi(t) \rangle_\rho .
\label{selfconsistent}
\end{equation}
Here $\langle \cdot \rangle_{\rho }$ stands for the statistical average
with respect to the probability density $\rho$ associated with
Eq.~(\ref{meanfieldone}) or Eq.~(\ref{meanfieldtwo}).

The Fokker-Planck equation for the probability density $\rho(\varphi, t;
\varphi_0(t))$ in the one-field case follows immediately from
the Langevin equation~(\ref{meanfieldone}),
\begin{equation}
\frac{\partial}{\partial t} \rho =
- \frac{\partial}{\partial \varphi} \left[a(\varphi; \varphi_0(t))
\rho\right] +
\frac{\sigma^2}{2} \frac{ \partial ^2}{\partial \varphi^2} \left[
\Gamma(\varphi) \rho \right] .
\label{FP1}
\end{equation}
We explicitly note the dependence of $\rho$ on the
unknown mean field $\varphi_0(t)$, which must be determined via
Eq.~(\ref{selfconsistent}) using the solution of the Fokker-Planck
equation.  The time dependent probability density has not
been found analytically.  In the stationary state the left side
of Eq.~(\ref{FP1}) is set to zero and the equation can be solved, to
yield the steady state probability density
\begin{eqnarray}
\rho_{\mathrm{st}}(\varphi;\varphi_0) &=& N(\varphi_0)
\Gamma(\varphi)^{-1} e^{-(2/\sigma^2)\left[ V(\varphi)
+\frac{K}{2}(\varphi_0 -\varphi)^2\right]}\nonumber\\
&=& N(\varphi_0) e^{-(2/\sigma^2)V_{eff}(\varphi)}, \label{stat}
\end{eqnarray}
where
\begin{equation}
V_{eff}(\varphi)\equiv V(\varphi) +\frac{K}{2}(\varphi_0
-\varphi)^2 +\frac{\sigma^2}{2}\ln \Gamma(\varphi) \label{effpot}
\end{equation}
and $N(\varphi_0)$ is the normalization constant.  The mean field
$\varphi_0$ can then be found from Eq.~(\ref{selfconsistent}). A
disordered stationary phase is associated with the solution
$\varphi_0=0$, while a solution $\varphi_0\neq 0$ corresponds to
an ordered stationary phase. The phase boundaries for different
forms of $V(\varphi)$ and $\Gamma(\varphi)$ are detailed
in~\cite{buceta04}. In particular, it shows that even if the
potential $V(\varphi)$ is monostable (single-well), the presence
of field-dependent multiplicative noise leads to an effective
bistable (double-well) potential in certain regions of the
parameter space. This is  precisely the manifestation of a
noise-induced phase transition. More elaborate forms of
$\Gamma(\varphi)$ could even lead to multistable potentials. Note
that the procedure in~\cite{buceta04} leads to a complete mean
field stationary state analysis, but does not provide information
about the dynamics of the approach to the steady state.

For the two-field case the Fokker-Planck equation for the probability
density $\rho(\varphi, z, t; \varphi_0(t),z_0(t))$ follows from
the Langevin equation~(\ref{meanfieldtwo}),
\begin{eqnarray}
\frac{\partial}{\partial t} \rho &=&
- \frac{\partial}{\partial \varphi} \left [\left(a(\varphi; \varphi_0(t))
-\omega z\right)\rho\right]\nonumber\\
&&- \omega \varphi \frac{\partial}{\partial
z} \rho +
\frac{\sigma^2}{2} \frac{ \partial ^2}{\partial \varphi^2} \left[
\Gamma(\varphi) \rho \right] .
\label{FP2}
\end{eqnarray}
This equation has not been solved analytically.  Furthermore, since we
know from numerical simulations~\cite{buceta06} that the system supports
collective oscillations, it is necessary to solve the time-dependent
problem even to find the long-time behavior.  The only available
information to date is numerical.

To obtain time-dependent solutions to the one-field and two-field
models, we will assume a Gaussian form for the evolving probability
density with time-dependent parameters to be found self-consistently
from the associated Fokker-Planck equation. Thus, for the one-field
problem we take
\begin{equation}
\rho(\varphi,t) = e^{A(\varphi-\varphi_0)^2+C},
\label{gaussianone}
\end{equation}
where the time dependent mean field $\varphi_0(t)$ and inverse width
parameter $A(t)$ are to be determined.
The parameter $C(t)$ is found from the normalization condition
\begin{equation}
\int d\varphi \rho(\varphi) =1
\end{equation}
to be given by
\begin{equation}
C(t) = \frac{1}{2} \ln\left( \frac{-A(t)}{\pi} \right).
\label{C1}
\end{equation}
It is immediately evident from a comparison of Eqs.~(\ref{gaussianone})
and (\ref{stat}) that (\ref{gaussianone}) is at best an
approximate solution, but we will subsequently show that for
sufficiently strong coupling $K$ this in fact provides an excellent
approximation for the mean value and width of the distribution.
Note that Eq.~(\ref{stat}) implies that in the mean field approximation
$ A = - V_{eff}''(\varphi_0)/\sigma^2 $ for the stationary state.
For the two-field case we posit the form
\begin{equation}
\rho(\varphi,z,t) = e^{A(\varphi-\varphi_0)^2+E(\varphi-
\varphi_0)(z-z_0) + M(z-z_0)^2+C},
\label{gaussiantwo}
\end{equation}
where now $z_0(t)$, $E(t)$ and $M(t)$ are also to be determined.
The normalization condition
\begin{equation}
\int d\varphi dz \rho(\varphi,z,t) =1
\end{equation}
fixes $C(t)$,
\begin{equation}
\label{C2}
C = \frac{1}{2} \ln\left( \frac{4MA-E^2}{4\pi^2} \right) .
\end{equation}
Note that the Gaussian only has finite norm for $4MA-E^2>0$.

We expect these forms to work best for strong coupling.  If
coupling is too weak, then there is no transition to collective
behavior and a Gaussian ansatz is not appropriate.  As coupling
becomes extremely strong, the distribution approaches a
$\delta$-function.  This provides the motivation for a narrow
distribution, whose mean and width we assume to be well captured
by a Gaussian when coupling is strong.   The Gaussian ansatz thus
rests on the observation that $\varphi_0(t) \sim z_0(t) \sim
A\langle (\delta \varphi)^2\rangle \sim E\langle \delta \varphi
\delta z\rangle \sim M \langle (\delta z)^2\rangle \sim O(K^0)$
together with the assumption that the second order moments
$\langle (\delta \varphi)^2\rangle \sim \langle (\delta z)^2\rangle
\sim O(1/K)$ and the expectation that higher order moments are
subleading for large $K$.  Here $\delta \varphi \equiv
\varphi-\varphi_0$ and $\delta z \equiv z-z_0$.
In the following sections we show that these assumptions lead to
consistent results.

\section{Single field}
\label{sec4}

We start with Eqs.~(\ref{gaussianone}) and (\ref{C1}), substitute
the Gaussian into the Fokker-Planck equation~(\ref{FP1}), Taylor
expand $a$, $\Gamma$, and their derivatives, and implement a
large-$K$ expansion considering the previous comments. We also
recognize that the kinetic coefficient and its derivatives, as
well as the local potential and its derivatives, are independent
of $K$. The contributions to leading orders in $K$ then result in
the set of equations
\begin{eqnarray}
\dot A &=& 2 A \Gamma_0 ( K + \sigma^2 A ), \label{dotAlargeK} \\
\dot \varphi_0 &=& -\Gamma_0 V_0' - \left( \frac{3\sigma^2}{2} +
  \frac{K}{A} \right) \Gamma_0', \label{dotphi0largeK}
\end{eqnarray}
For each function $f(\varphi)$ we have adopted the notation
$f_0\equiv f(\varphi_0)$, $f'_0 \equiv
[df(\varphi)/d\varphi]_{\varphi=\varphi_0}$, etc.  Note that it is
clear from these two equations that the system evolves towards
$A\sim O(K)$ and $\varphi_0\sim O(1)$. We also point out that the
normalization condition~(\ref{C1}) is consistent with the
evolution equation one obtains for $C(t)$, namely, $\dot C
=\Gamma_0 ( K + \sigma^2 A )$. The set of equations
(\ref{dotAlargeK}) and (\ref{dotphi0largeK}) is of course
nonlinear and can not be solved exactly analytically, but it is
merely a set of two ordinary differential equations whose
numerical solution is trivial.

Even so, we can make further analytic progress by noting that for
large $K$, and provided the initial value of $A$ is of (negative) $O(K)$,
the evolution of $A$ toward its stationary state is clearly faster than
that of $\varphi_0$, thus allowing us to consider $\Gamma_0$ as nearly
constant during the relaxation of $A$.  In this approximation
Eq.~(\ref{dotAlargeK}) is a Riccati equation with solution
\begin{equation}
\label{riccati}
A(t) = \frac{-K}{\sigma^2} \frac{1}{1- \left(1+\frac{K}{\sigma^2
A(0)}\right) e^{-2 \Gamma_0 K t}}.
\end{equation}
After the stationary state for $A$ is reached, the evolution of
$\varphi_0(t)$ is governed by
\begin{equation}
\label{uncoup}
\dot \varphi_0 =  - \Gamma_0 \frac{d V_{eff}(\varphi_0)}{d\varphi_0}.
\end{equation}

Before comparing the results of the theory with those of numerical
simulations,
we note that in the stationary state where $\dot A = \dot \varphi =0$
we find from Eq.~(\ref{dotAlargeK}) or Eq.~(\ref{riccati})
that $A=-K/\sigma^2$, and that $\varphi_0$ obtained
from Eq.~(\ref{dotphi0largeK}) or (\ref{uncoup}) is
the solution of the condition $dV_{eff}(\varphi_0)/d\varphi_0=0$.
The latter is exactly the mean field solution to the problem, which
is thus recovered from the Gaussian ansatz. The former differs from the
exact inverse width of the distribution (\ref{stat}) by
contributions of $O(1)$.

\begin{figure}
\includegraphics[width=8 cm]{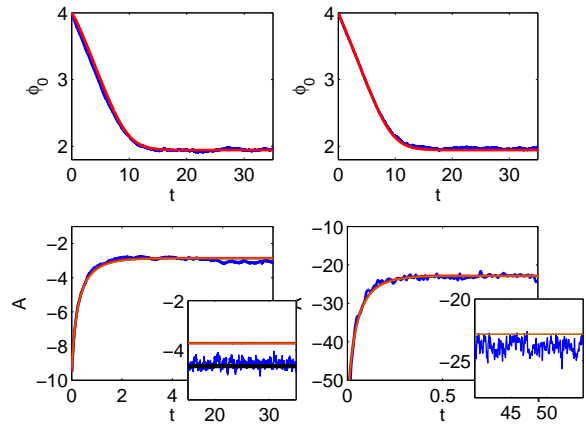}
\caption{(Color online) The time evolution of $\varphi_0$ (top
panels) and $A$ (bottom panels), that give the mean and the
inverse width of the distribution, is shown for simulations of
\emph{globally coupled} arrays of $N=4000$ sites and compared to
the numerical solution of Eqs.~(\ref{dotAlargeK}) and
(\ref{dotphi0largeK}). The left column represents modest coupling,
$K=10$, while the right column shows $K=80$; $\sigma^2=3.5$ for
all plots. The initial values $-A(0)$ are chosen to be $\gtrsim
K/\sigma^2$; specifically, $A(0)=-9.7$ in the left column and
$A(0)=-58$ in the right column. The dark (blue) lines show data
from lattice simulations, while the light (red) lines represent
theoretical predictions.  The very light (brown) curves in the
lower right panel and inset represent the uncoupled dynamics given
by Eq.~(\ref{uncoup}) and the Riccati equation~(\ref{riccati}).
The lower left inset shows the simulation results and the
prediction of the mean field theory (solid black line) in the
stationary state, which is exact when $N\to\infty$.  The inset in
the bottom right figure shows a close up of the late time
evolution.  Note the different horizontal and vertical scales in
the various panels.} \label{SimTh2}
\end{figure}

However, the ansatz takes us beyond the stationary solution to provide
information about the dynamics of the system as it approaches the steady
state. In Fig.~\ref{SimTh2} we show four sets of results for the mean
field $\varphi_0(t)$ and the inverse width parameter $A(t)$.  One
is the outcome
of the direct simulation of globally coupled arrays for moderate and for
strong coupling $K$. The second is the result of integrating
Eqs.~(\ref{dotAlargeK}) and (\ref{dotphi0largeK}), the third is the
outcome of Eqs.~(\ref{riccati}) and (\ref{uncoup}) that assume different
relaxation rates, and the fourth is the outcome of the mean field
distribution (\ref{stat}).  In these and all subsequent figures we have
made the representative choices
\begin{equation}
V(\varphi)=\frac{\varphi^2}{2}, \quad \Gamma(\varphi) =
\frac{1+\varphi^2}{1+\varphi^4},
\label{choices}
\end{equation}
which were also used in our earlier work~\cite{buceta04,wood06,buceta06}.

The first conclusion is that
the time scale of relaxation and the steady state value of
$\varphi_0$ are correctly predicted by the large coupling theory, even for
modest values of $K$.  The agreement is spectacular during the approach
to steady state.  In addition, we note that the time scale separation
between $A$ and $\varphi_0$ is quite evident for large $K$ and is
a reasonable assumption also for modest values, $K \approx 10$.
Further discussion of the results in this figure requires that we take
note of the different horizontal and vertical scales in the different
panels.  A small but consistent discrepancy between the Gaussian theory
results for $A$ and simulation results for the inverse width of the
distribution arises at long times owing to neglected
higher order contributions in the theory.  In the lower
left panel we have expanded the vertical scale to make the difference
clear, but note that it is extremely small for large $K$ (see lower
right inset).  The lower left inset confirms that the simulation and
mean field theory stationary widths are in fact identical, as they
should be for sufficiently large arrays.
While we only show results for two values of $K$ and one
lattice size, the results for a range of values of $K$ ($K=20$, $40$,
$60$), of lattice sizes ($N = 250$ up to $N = 16000$), and of initial
conditions in which $A(t=0) \sim -K$ follow the patterns described
above.  We thus conclude that the Gaussian ansatz theory gives
quantitatively excellent results for the evolution of the globally coupled
system toward the steady state and for the steady state itself when the
coupling is strong and the width of the initial distribution is of the
same order as that of the steady state.

\begin{figure}
\begin{center}
\includegraphics[width=8 cm]{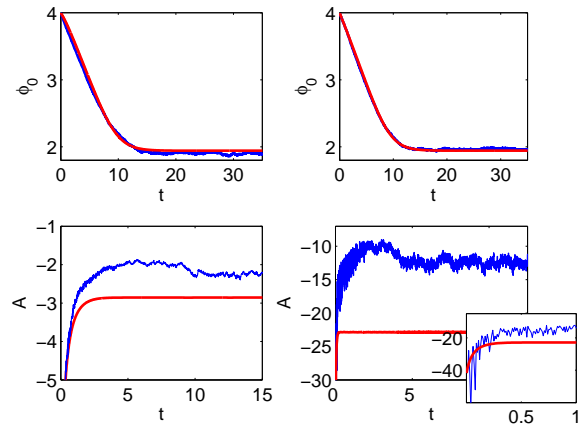}
\end{center}
\caption{(Color online) The time evolution of the mean (top
panels) and inverse width (bottom panels) is shown for simulations
of \emph{locally coupled} two-dimensional arrays of $N=64\times
64$ sites and compared to the numerical solution of
Eqs.~(\ref{dotAlargeK}) and (\ref{dotphi0largeK}). The left column
represents modest coupling, $K=10$, while the right column shows
results for $K=80$; $\sigma^2=3.5$ for all plots. The dark (blue)
lines show data from lattice simulations, while the light (red)
lines represent theoretical predictions. The initial values are
$A(0)=-9.2$ (left column) and $-42$ (right column). The inset in
the bottom right panel shows a close up of the early time
evolution.} \label{SimTh1}
\end{figure}

Our theory is based on a mean field theory, and so the appropriate
comparison with
simulations is as we have shown in Fig.~\ref{SimTh2}, with a globally
coupled array.  However, mean field theories are often used to describe
locally coupled systems, and so we compare our theoretical results with
simulation results in which the units in a two-dimensional array
are connected only to their nearest neighbors. The results are shown in
Fig.~\ref{SimTh1}.
The theory accurately captures the behavior of $\varphi_0$, including
both the transient and steady state dynamics, in the large and modest
$K$ regimes.  However, the inverse width of the distribution
(as given indirectly
by $A$) is underestimated by the theory, as it is by the original mean
field theory in this model.  A discrepancy of this sort is a ubiquitous
feature of mean field theories, which are principally designed to capture
the mean field value.

\begin{figure}
\begin{center}
\includegraphics[width=8 cm]{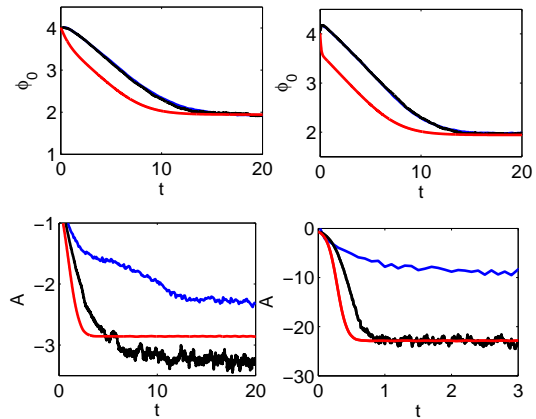}
\end{center}
\caption{(Color online) The time evolution of the mean field (top
panels) and the inverse width of the distribution (bottom panels)
is shown for both locally (dark or blue) and globally (black)
coupled simulations and compared to the numerical solution (light
or red) of the large $K$ theory.  Again, the left column is for
$K=10$ and the right for $K=80$. Here we choose $|A(0)|<1 $
(specifically, $A(0)\sim -0.1$), and the transient dynamics are
not well-described for early times. The theory accurately captures
the steady-state behavior of the globally coupled simulations in
the large $K$ regime. $N=4000$ (globally coupled) and $64\times
64$ (locally coupled), and $\sigma^2=3.5$ for all panels.}
\label{SimTh3}
\end{figure}

The comparisons so far have relied on the initial value $A(t=0)$ being
of (negative) $O(K)$, that is, an initial distribution whose width is
of the same order as that of the steady state.
When the initial width of the distribution is much larger than that of
the steady state, i.e., when $A(t=0)$ is very different from (much
smaller in magnitude than) $K$, it becomes more problematic to
capture the transient
dynamics, although the steady state behavior is still predicted
accurately.  This is shown in Fig.~\ref{SimTh3}.

\begin{figure}
\begin{center}
\includegraphics[width=8 cm]{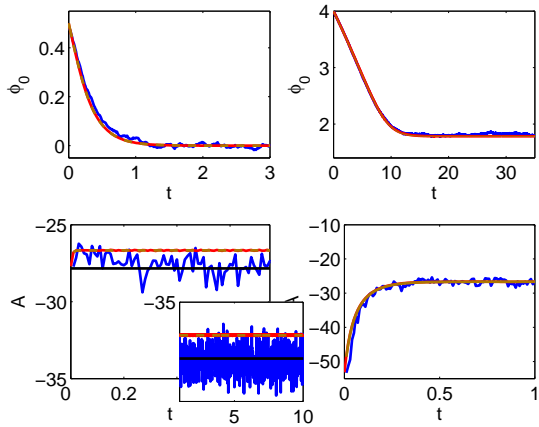}
\end{center}
\caption{(Color online) Multistability is captured by the large
$K$ theory.  The left panel shows the dynamics of the mean field
and the inverse width of the distribution for initial conditions
leading to a disordered state ($\varphi_0(0) \approx 0.5$,
$A(0)=-29$).  The right panel shows the corresponding plots for
initial conditions leading to an ordered phase
($\varphi_0(0)\approx 4$, $A(0)=-55$).  The inset in the bottom
left shows a close up of the steady state behavior.  Dark (blue):
globally coupled simulations; light (brown): Gaussian ansatz
theory; black: mean field theory. In all plots, $K=80$,
$\sigma^2=3.0$.} \label{mult1}
\end{figure}

Finally and importantly, we note that the large coupling
theory accurately captures the multistable nature of the dynamics
(Fig.~\ref{mult1}).  This is a new feature of this theory that
provides dynamical information not provided by
the usual mean field theory.  In our earlier work we had established this
multistability only through direct numerical simulations of the array.
For large $K$, the agreement between theory and
simulations both in the dynamical regime and in the steady state is
quite remarkable for initial conditions leading to the
ordered state (right panel).  While the mean field theory of course
predicts the width of the distribution exactly, the Gaussian ansatz
slightly overestimates the width for initial conditions leading to a
disordered phase (left panel).  The disordered state is marked by a
relatively broad distribution, and, not
surprisingly, given the underlying requirements stated earlier, the
theory does not exactly capture this prediction.

\section{Two fields}
\label{sec5}

The more stringent test of the theory lies in the two-field
system, where there are only numerical simulation results. Now we
begin with Eqs.~(\ref{gaussiantwo}) and (\ref{C2}), substitute the
assumed distribution into the two-variable Fokker-Planck equation
(\ref{FP2}), and again implement a large-$K$ expansion. This
results in the set of equations for the coefficients in the
Gaussian \footnote{ Note that if we consider $A \sim M \sim E \sim
K$ at leading order, the terms containing $\omega$ will not be
retained in Eq.~\eqref{ameDyn}. However, at the stationary point
$E=0$, so it is interesting to also consider the case when $E\sim
K^0$ and keep the additional terms that are relevant in this case.
Therefore, in Eq.~\eqref{ameDyn} we have considered the
possibility that $A \sim M \sim K$ and $E \sim K^0$, and have kept
the leading order terms for this case.},
\begin{equation}
\label{ameDyn}
\begin{split}
\dot A &= 2A\Gamma_0 (K+\sigma^2 A), \\
\dot M &= \left( \frac{\sigma^2}{2}\Gamma_0 E + \omega \right) E, \\
\dot E &= \Gamma_0 (K+2 \sigma^2 A) E + 2\omega (A-M),
\end{split}
\end{equation}
along with those for the mean values
\begin{equation}
\label{evolPhiLargeK}
\begin{aligned}
\dot \varphi_0 &= - \Gamma_0 V_0'-\left(\frac{3\sigma^2}{2}
+ \frac{K}{A} \right) \Gamma_0' -
\omega z_0, \\
 \dot z_0 &= \omega \varphi_0.
\end{aligned}
\end{equation}
Again, it is easy to ascertain that the normalization condition
(\ref{C2}) is consistent with the evolution equation obtained for
$C(t)$.

The solution we are interested in is a collective oscillatory
mode, which of course requires the time-dependent solution of the
coupled sets of ordinary differential equations (\ref{ameDyn}) and
(\ref{evolPhiLargeK}), a task which is vastly simpler than the
solution of the time-dependent Fokker-Planck equation.  However,
as in the single field case, we can further simplify the problem
of finding the oscillatory long-time behavior of the mean values
$\varphi_0$ and $z_0$ by exploring the regime where the
coefficients $A$, $M$, and $E$ have reached a steady state, that
is, by setting the left hand sides in Eq.~(\ref{ameDyn}) equal to
zero.  There are four stationary solutions, one of which is
\begin{equation}
\label{stE0}
A = \frac{-K}{\sigma^2}, \quad E = 0, \quad M = A.
\end{equation}
The other solutions, $ (A,\, E,\, M) = (0,\, 0,\, 0) $, $ (A,\, E,\, M) =
(-K/\sigma^2,\, -2\omega/(\sigma^2\Gamma_0),\, 0) $, and $ (A,\, E,\, M) = (0,
\, -2\omega/(\sigma^2\Gamma_0),\, -K/\sigma^2) $
do not satisfy the condition $4MA-E^2>0$ necessary for proper
normalization, cf. Eq.~(\ref{C2}), and are hence unphysical.

For the long-time behavior, it now remains to substitute the stationary
$A$ into Eq.~(\ref{evolPhiLargeK}) and solve the coupled set of just two
equations,
\begin{equation}
\label{evolPhiLargeKlate}
\begin{aligned}
\dot \varphi_0 &= - \Gamma_0 \frac{d
V_{eff}(\varphi_0)}{d\varphi_0} -
\omega z_0, \\
 \dot z_0 &= \omega \varphi_0.
\end{aligned}
\end{equation}
We note that this set has one stationary solution, the disordered
state $(\varphi_0,z_0)=(0,0)$.

We can go even further toward the analytic oscillatory solution by
implementing a multiscale perturbation theory~\cite{multiscale}.
For this purpose we combine (\ref{evolPhiLargeK}) into a single
second-order differential equation,
\begin{equation}
\label{eff}
\begin{split}
&\ddot z + \omega^2 z = \omega G(\varphi), \\
&G(\varphi) = - \Gamma(\varphi) \frac{d V_{eff}(\varphi)}
{d \varphi} ,
\end{split}
\end{equation}
with $ \varphi \equiv \dot z/\omega$.  It is understood that
the variables are the mean fields and should therefore carry the $0$
subscript, which we have omitted for economy of notation.  We treat
$G$ as a perturbation and write
\begin{equation}
\ddot z + \omega^2 z = \epsilon \omega G(\dot z / \omega),
\end{equation}
where $\epsilon$ is a small parameter.  The solution is then expressed
in terms of different time scales,
($ T_0 $, $ T_1 $, \ldots),
\begin{equation}
z(t) = Z(T_0,T_1,\cdots) = Z_0(T_0,T_1) + \epsilon Z_1(T_0,T_1) +
O(\epsilon^2),
\end{equation}
where $T_n \sim O(\epsilon^n T_0)$.  For the time derivatives we have
\begin{equation}
\begin{aligned}
\frac{d z}{d t} &= \frac{\partial Z}{\partial T_0} +
  \epsilon \frac{\partial Z}{\partial T_1} =
  \frac{\partial Z_0}{\partial T_0} + O(\epsilon), \\
\frac{d^2 z}{d t^2} &= \frac{\partial^2 Z}{\partial T_0^2} +
\epsilon \frac{\partial^2 Z}{\partial T_0 \partial T_1} +
O(\epsilon^2) \\ &= \frac{\partial^2 Z_0}{\partial T_0^2} + \epsilon
\frac{\partial^2 Z_1}{\partial T_0^2} + \epsilon \frac{\partial^2
Z_0}{\partial T_0 \partial T_1} + O(\epsilon^2)
\end{aligned}
\end{equation}
The evolution equation at zero order in $ \epsilon $ is
\begin{equation}
\frac{\partial^2 Z_0}{\partial T_0^2} + \omega^2 Z_0 = 0,
\end{equation}
which has a solution
\begin{equation}
\label{sol0}
Z_0(T_0,T_1) = R(T_1) \cos(\omega T_0) + S(T_1) \sin(\omega T_0).
\end{equation}

The coefficients $R(T_1)$ and $S(T_1)$ are determined by considering the
evolution equations at first order in $\epsilon$,
\begin{equation} \label{EvolEqEpsilon1}
\frac{\partial^2 Z_1}{\partial T_0^2} + \omega^2 Z_1 = -
\frac{\partial^2 Z_0}{\partial T_0 \partial T_1} + \omega\,
G\!\left( \frac{1}{\omega} \frac{\partial Z_0}{\partial T_0}
\right).
\end{equation}
As always in multiscale perturbation theory,
in order to avoid secular terms in the solution
$R$ and $S$ must be chosen in such a way that resonant terms do
not appear in the right hand side of Eq.~\eqref{EvolEqEpsilon1},
i.e.,
\begin{equation}
\begin{split}
&\int_{-\infty}^{\infty}{dT_0 \, \cos(\omega T_0) \left[ -
  \frac{\partial^2 Z_0}{\partial T_0 \partial T_1} + \omega\,
  G\!\left( \frac{1}{\omega} \frac{\partial Z_0}{\partial T_0}
  \right) \right]} = 0, \\
&\int_{-\infty}^{\infty}{dT_0 \, \sin(\omega T_0) \left[ -
  \frac{\partial^2 Z_0}{\partial T_0 \partial T_1} + \omega\,
  G\!\left( \frac{1}{\omega} \frac{\partial Z_0}{\partial T_0}
  \right) \right]} = 0,
\end{split}
\end{equation}
where from (\ref{sol0}) it follows that
\begin{equation}
\frac{\partial^2 Z_0}{\partial T_0 \partial T_1} = -
\frac{\partial R}{\partial T_1} \;\omega \sin(\omega T_0) +
\frac{\partial S}{\partial T_1} \;\omega \cos(\omega T_0).
\end{equation}
This immediately leads to the equations for $ R $ and $ S $,
\begin{equation}
\label{ABevol}
\begin{split}
&\frac{\partial R}{\partial T_1} = - \frac{
 \int_{-\infty}^{\infty}{dT_0 \, \sin(\omega T_0) G\!\left(
 \frac{1}{\omega} \frac{\partial Z_0}{\partial T_0} \right) } } {
 \int_{-\infty}^{\infty}{dT_0 \, \sin^2(\omega T_0) } }, \\ \\
&\frac{\partial S}{\partial T_1} = \frac{
 \int_{-\infty}^{\infty}{dT_0 \, \cos(\omega T_0) G\!\left(
 \frac{1}{\omega} \frac{\partial Z_0}{\partial T_0} \right) } } {
 \int_{-\infty}^{\infty}{dT_0 \, \cos^2(\omega T_0) } }.
\end{split}
\end{equation}
The quotients can be computed by introducing a cutoff $ \Lambda $ in
the integrals, $ \int_{-\Lambda}^\Lambda dT_0 \cdots$, and
subsequently taking the limit $ \Lambda \to \infty $.

We thus have theoretical predictions at three levels of
approximation. The most detailed is the five-equation set
(\ref{ameDyn}) and (\ref{evolPhiLargeK}). This set of equations
contains the dynamical approach to the long time behavior.  The
second level is contained in Eqs.~(\ref{stE0}) and
(\ref{evolPhiLargeKlate}). This yields the long-time oscillatory
behavior of the mean field and the width parameters of the
distribution at long times. Finally, Eq.~(\ref{stE0}) together
with (\ref{EvolEqEpsilon1}) and (\ref{ABevol}) (along with the
second equation in (\ref{evolPhiLargeK})) provide a full long-time
solution in terms of quadrature.  Note that this explicit solution
presents a circular limit cycle of the form $ z = S_0 \sin(\omega t) $
with $ S_0 $ constant [$ Z_0 = S_0 \sin(\omega T_0) $ ], provided we have $
\partial R/\partial T_1 =  0 $ [directly satisfied for the choice in
(\ref{choices})] and $ \partial S/\partial T_1 =  0 $, i.e., if
\begin{equation}
\label{circ}
0 = I(S_0)=\int_{-\infty}^{\infty}{dT_0 \, \cos(\omega T_0) \, G\!\left(
S_0 \cos(\omega T_0) \right)}.
\end{equation}
Since our numerical simulations~\cite{buceta06} indicate an essentially
circular limit cycle near the onset of multistability, we expect that
Eq.~(\ref{circ}) may provide an accurate prediction of the onset in the
strong coupling limit.

We now proceed to test our multi-level theoretical predictions
against direct numerical simulations of the array.  Again, we adopt the
representative choices (\ref{choices}) for the potential and relaxational
functions.  We concentrate on the multistability onset regime by fixing
the value of the noise intensity appropriately.

\begin{figure}
\includegraphics[width=8 cm]{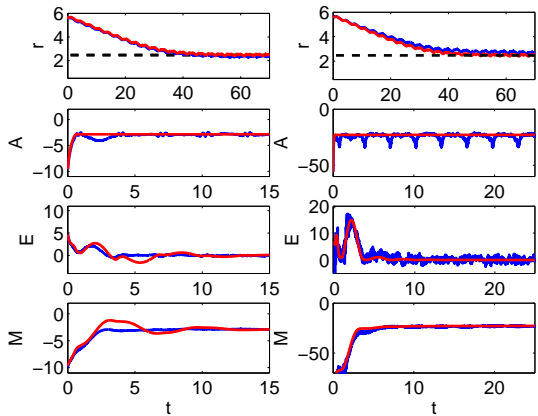}
\caption{(Color online) The time evolution of the limit cycle
radius, $r$ (top panels) and of the Gaussian ansatz coefficients
(three subsequent panels). The dark (blue) curves are appropriate
moment results from simulations of globally coupled arrays
($N=4000$ sites), and the light (red) curves are obtained from the
solution of Eqs.~(\ref{ameDyn}) and (\ref{evolPhiLargeK}). $K=10$
in the left column and $K=80$ in the right column. Initial values
for left column: $(A,E,M)=(-9.6,4.7,-9.7)$; right column:
$(A,E,M)=(-56,7,-79)$.  The noise intensity $\sigma^2=3.5$ in all
panels. The dashed line in the top panels is the steady state
radius predicted by multiscale analysis. } \label{2dof1b}
\end{figure}

\begin{figure}
\includegraphics[width=8 cm]{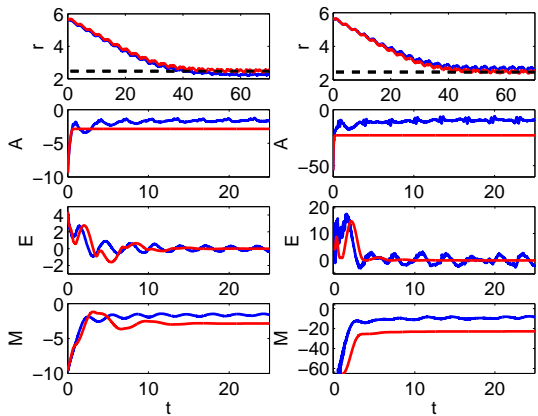}
\caption{(Color online) Same as Fig.~\ref{2dof1b}, but now the
simulations are for locally coupled arrays. Initial values for
left column: $(A,E,M)=(-9.4,4.4,-9.6)$; right column:
$(A,E,M)=(-55,5,-76)$. } \label{2dof1a}
\end{figure}

\begin{figure}
\includegraphics[width=8 cm]{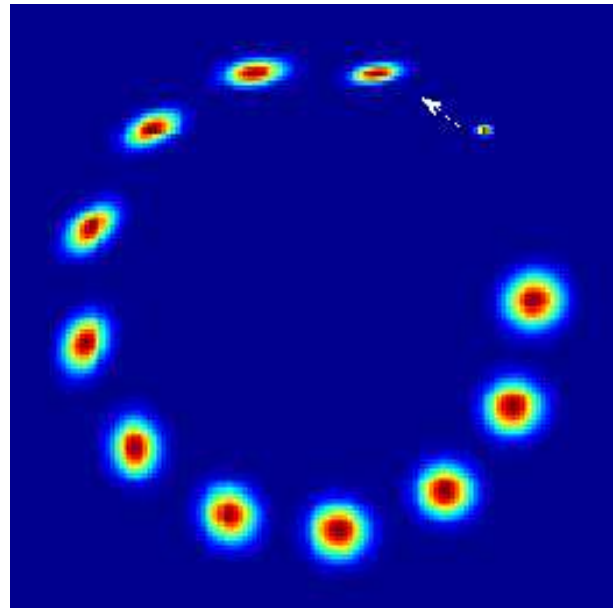}
\caption{(Color online) Evolution of the theoretical probability
distribution function as obtained by substituting the numerical
solution of Eqs.~(\ref{ameDyn}) and (\ref{evolPhiLargeK}) into
Eq.~(\ref{gaussiantwo}). The distribution becomes symmetrical
(circular) in $\varphi_0$ and $z_0$ prior to reaching its steady
state radius.  The evolution is only shown for the short time
leading up to relaxation of the distribution shape.  Over longer
times, the distribution will continue a circular trajectory whose
radius eventually reaches its steady state value (see
Fig.~\ref{2dof2Vers2}).  $K=10$ and $\sigma^2=3.5$.}
\label{pdfEvolv}
\end{figure}

Figure~\ref{2dof1b} compares the results of simulations for
globally coupled arrays with those of our theory.
The agreement is clearly excellent for the limit cycle radius as
well as the width parameters for all times. The time scale
separation required for the validity of the large $K$ equations is
clearly satisfied for $K\sim 80$ and even for modest values,
$K\sim 10$.  Note that the width parameters reach
the steady state very quickly even when $E$ is initially chosen to
be of $O(K)$.  The theory very slightly overestimates the steady
state radius $ r \equiv \sqrt{\dot z^2 / (2\omega^2) + z^2/2} $
for modest $K=10$, and very slightly underestimates the radius for
large $K=80$ (neither visible on the
scale of the figures). Again as expected, the agreement with the
simulations for the locally coupled array is less spectacular
(Fig.~\ref{2dof1a}), but the limit cycle radius is still captured
very accurately for all times. For visualization purposes, we also
show the early time evolution of the theoretical probability
distribution function in Fig.~\ref{pdfEvolv}.

\begin{figure}
\includegraphics[width=8 cm]{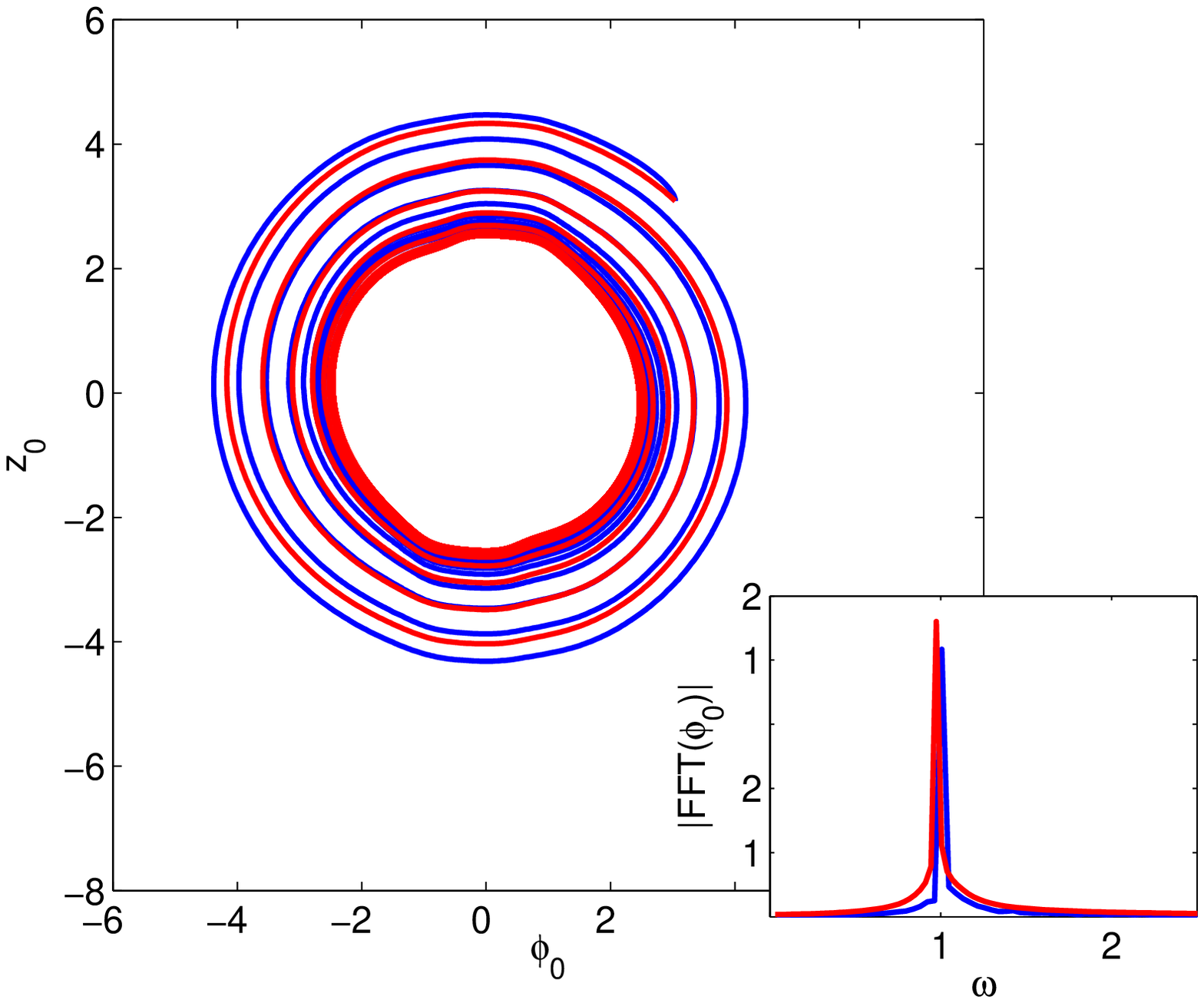}
\caption{(Color online) The phase space portraits of lattice
simulations (dark or blue) and the large $K$ equations
(\ref{ameDyn}) and (\ref{evolPhiLargeK}) (light or red) show good
agreement in terms of the steady state limit cycle radius, though
the theory slights underestimates $r$. Inset: oscillation
frequency.  $K=80$, $\sigma^2=3.5$, and $\omega=1$.}
\label{2dof2Vers2}
\end{figure}

Figure~\ref{2dof2Vers2} confirms that the phase portrait of the
limit cycle from the globally coupled lattice simulations
corresponds reasonably well to that predicted by the large $K$
equations, and that the frequency of the oscillations is also
accurately predicted.  The dynamic evolutions of the simulation and
of the theory do not match exactly because the small error in
the frequency implies an increasing dephasing with time. The
figure shows the case $K=80$, but the results are only
representative, and again the theory holds well even for more
modest values of $K$ and for a range of initial conditions leading
to limit cycle behavior.

\begin{figure}
\includegraphics[width=8 cm]{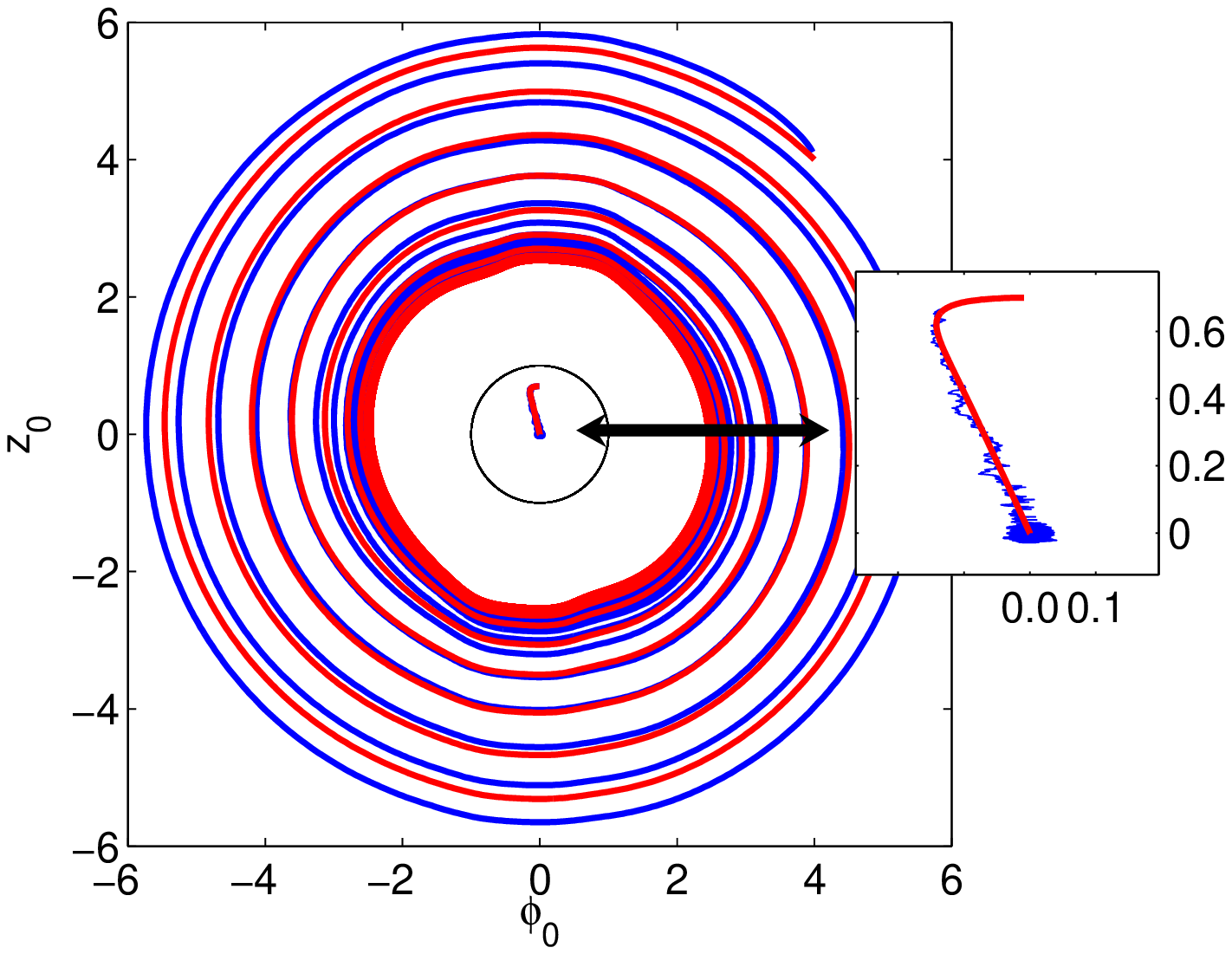}
\caption{(Color online) Comparison of the phase space portraits of
lattice simulations (dark or blue) and the large $K$ equations
(light or red) show that the latter correctly predict multistable
behavior.  The right inset shows a close up of the indicated
portion of the phase portrait. The black curve is the unstable
limit cycle obtained from the multiscale analysis. $K=80$,
$\sigma^2=3.5$, and $\omega=1$.} \label{multstab2}
\end{figure}

\begin{figure}
\includegraphics[width=8 cm]{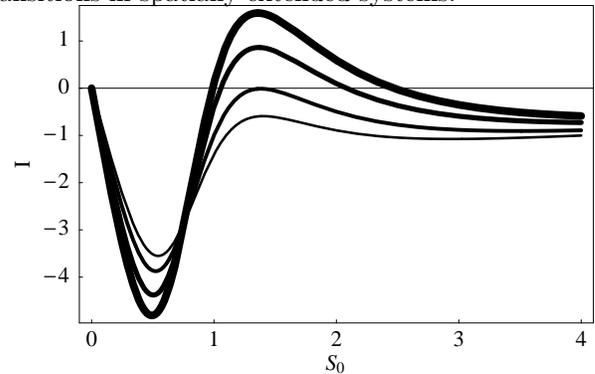}
\caption{The multiscale analysis predicts the onset of bistability
to occur at $\sigma^2 \approx 2.4$.  Curves are, from thickest to
thinnest, $\sigma^2=3.5$, $3.0$, $2.4$, $2.0$.}
\label{MultScaleFig}
\end{figure}

Finally, as in the single field problem, the large coupling theory
predicts the occurrence of multistable regions in parameter space.
In our earlier work~\cite{buceta06} we noted the expectation of
multistability, but did not explicitly pursue it in our
simulations, nor did we have a predictive theory as we now do.  In
Fig.~\ref{multstab2} we show that the multistable nature of the
phase transition is fully captured by the large coupling theory.
In particular, different initial conditions lead to either a
disordered phase or a limit cycle. To test whether the theory
correctly predicts the onset of multistability is more cumbersome,
but we can at least do it easily in the multiscaling (large $K$)
regime on the basis of Eq.~(\ref{circ}). The value of the noise
for onset of multistability in general depends on coupling
strength~\cite{buceta04,buceta06}, but this dependence greatly
weakens with increasing $K$.  For example, our simulation results
indicate that at $K=10$ multistability first occurs at roughly
$\sigma^2 \approx 3$, with the noise value decreasing ever more
slowly with increasing $K$ (e.g., $\sigma^2 \approx 2.6$ for
$K=20$, $\sigma^2 \approx 2.5$ for $K=40$).  At $K=80$ the
transition occurs at $\sigma^2 \approx 2.4$. Equation~(\ref{circ})
is in fact independent of $K$. The function $I(S_0)$ is shown
explicitly in Fig.~\ref{MultScaleFig}.  The first zero is the
disordered state and the second first appears when $\sigma^2=2.4$.
As noise increases, an intermediate unstable solution also
appears, illustrated explicitly in Fig.~\ref{multstab2}. The
multiscaling result thus accurately predicts not only the radius
$r=S_0$ of the limit cycle but the noise for onset of
multistability when coupling is strong, and the increase in the
limit cycle radius with increasing noise strength.

\section{Conclusions}
\label{sec6}

We have presented an analytical theory for the dynamics of
relaxational systems with field dependent coefficients.  Based on
a Gaussian ansatz and an expansion about the mean field values, we
derive ordinary differential equations detailing the time evolution
of the field distribution means and widths.  In the limit of large
coupling, the Gaussian
ansatz equations provide a consistent, normalized approximation of
the relevant probability distributions which agree with numerical
lattice simulations. In particular, our method allows us to study
the dynamics of both one- and two-field relaxational systems, including
those with oscillatory collective states previously beyond the reach
of the static mean field theories. This also provides an analytic
access to initial-condition-dependent multistable regimes for
noise-induced phase transitions in spatially extended systems.

\section*{Acknowledgments}
This work was partially supported by the National Science
Foundation under Grant No. PHY-0354937, and by the Ministerio de
Educaci\'on y Ciencia (Spain) under Grants Nos. FIS2006-05895,
FIS2005-24376-E, and BFM2003-02547/FISI. F.J.C. also thanks the
Universidad Complutense de Madrid for support through the program
\emph{Profesores Complutenses en el Extranjero}, and the Department of
Chemistry and Biochemistry of the University of California San
Diego for their hospitality.


\begin{thebibliography}{99}

\bibitem{horsthemke}
W. Horsthemke and R. Lefever, \emph{Noise-Induced Transitions}
(Springer, Berlin, 1984).

\bibitem{garcia}
J. Garc\'{\i}a-Ojalvo and J.M. Sancho. \emph{Noise in Spatially Extended
Systems} (Springer, New York, 1999).

\bibitem{chris}
C. Van den Broeck, J. M. R. Parrondo, and R. Toral, Phys. Rev. Lett.
{\bf 73}, 3395 (1994); C. Van den Broeck, J. M. R. Parrondo, R. Toral,
and R. Kawai, Phys. Rev. E {\bf 55}, 4084 (1997).

\bibitem{ibanes}
M. Iba\~nes, J. Garc\'{\i}a-Ojalvo, R. Toral, and J. M. Sancho, Phys.
Rev. Lett. {\bf 87}, 020601 (2001).

\bibitem{buceta04}
J. Buceta and K. Lindenberg,
%\emph{Comprehensive study of phase transitions in relaxational systems
%with field-dependent coefficients},
Phys. Rev. E {\bf 69}, 011102 (2004).

\bibitem{wood06}
K. Wood, J. Buceta, and K. Lindenberg,
%\emph{Comprehensive study of pattern formation in relaxational systems},
Phys. Rev. E {\bf 73}, 022101 (2006).

\bibitem{wio03} S. E. Mangioni and H. S. Wio, Phys. Rev. E {\bf 67},
056616 (2003).

\bibitem{wio04} B. von Haeften, G. Izus, S. Mangioni, A. D. Sanchez, and
H. S. Wio, Phys. Rev. E {\bf 69}, 021107 (2004).

\bibitem{kawai}
R. Kawai, X. Sailer, L. Schimansky-Geier, and C. Van den Broeck, Phys.
Rev. E {\bf 69}, 051104 (2004).

\bibitem{buceta06}
J. Buceta, K. Wood, and K. Lindenberg,
%\emph{Noise-induced
%oscillatory behavior in field-dependent relaxational dynamics},
Phys. Rev. E {\bf 73}, 042101 (2006).

\bibitem{qf}
S. Habib, Y. Kluger, E. Mottola, and J.P. Paz, Phys. Rev. Lett.
{\bf 76}, 4660 (1996); F. J. Cao and H. J. de Vega, Phys. Rev. D {\bf 63},
045021 (2001); D. Boyanovsky, F.J. Cao, and H. J. de Vega, Nucl.
Phys. B {\bf 632}, 121 (2002); F. J. Cao and H. J. de Vega, Phys. Rev. D
{\bf 65}, 045012 (2002).

\bibitem{multiscale}
L-Y. Chen, N. Goldenfeld, and Y. Oono,
%Renormalization group and singular perturbations: Multiple scales,
%boundary layers, and reductive perturbation theory
Phys. Rev. E {\bf 54}, 376 (1996).

\end{thebibliography}
\end{document}